\documentclass{webofc}
\usepackage[varg]{txfonts}   
\usepackage{cleveref}
\usepackage{mathtools}
\usepackage{xcolor}
\usepackage{wrapfig}
\usepackage[percent]{overpic}

\begin{document}
\title{Static quark anti-quark interactions at non-zero temperature from lattice QCD}
%
%

\author{\firstname{Gaurang} \lastname{Parkar}\inst{1}\fnsep\thanks{ \email{gaurang.parkar@uis.no}}
\and \firstname{Dibyendu} \lastname{Bala} \inst{2} 
\and \firstname{Olaf} \lastname{Kaczmarek} \inst{2} 
\and \firstname{Rasmus} \lastname{Larsen} \inst{1}
\and \firstname{Swagato} \lastname{Mukherjee} \inst{3}
\and \firstname{Peter} \lastname{Petreczky} \inst{3}
\and \firstname{Alexander} \lastname{Rothkopf} \inst{1}
\and \firstname{Johannes} \lastname{Heinrich Weber} \inst{4}
}

\institute{Faculty of Science and Technology, University of Stavanger, NO-4036 Stavanger, Norway \and
Fakult\"at f\"ur Physik, Universit\"at Bielefeld, D-33615 Bielefeld, Germany \and
Physics Department, Brookhaven National Laboratory, Upton, New York 11973, USA \and
Institut f\"ur Physik \& IRIS Adlershof, Humboldt-Universit\"at zu Berlin, D-12489 Berlin, Germany
}

\abstract{%
We present results on the in-medium interactions of static quark anti-quark pairs using realistic 2+1 HISQ flavor lattice QCD. Focus is put on the extraction of spectral information from Wilson line correlators in Coulomb gauge using four complementary methods. Our results indicate that on HISQ lattices, the position of the dominant spectral peak associated with the real-part of the interquark potential remains unaffected by temperature. This is in contrast to prior work in quenched QCD and we present follow up comparisons to newly generated quenched ensembles.
}
\maketitle
\section{Introduction}

The experimental investigation of heavy quarkonium in extreme conditions \cite{Rothkopf:2019ipj} is entering its next stage during run3 at the LHC at CERN. With major upgrades installed during the previous shutdown, the in-medium modification of the yields of both ground and excited state quarkonium will be measured in relativistic heavy-ion collisions \cite{Elfner:2022iae} with unprecedented precision. In addition new channels, such as the in-medium P-wave states \cite{Burnier:2016kqm} may finally be observed. Theory is urged to refine its understanding of in-medium heavy quarkonium from first principles in order to support its original mission, i.e. to act as a well controlled probe of the quark-gluon plasma. An important ingredient in the description of quarkonium dynamics in the past has been played by the complex in-medium potential (see e.g. \cite{Laine:2006ns}). It is an integral piece in the phenomenological modeling of quarkonium yields (see e.g. \cite{Islam:2020bnp,Katz:2015qja}) and has been connected to the in-medium quantum real-time dynamics of these bound states via the open-quantum systems framework \cite{Akamatsu:2020ypb}.

The notion of an interaction potential \cite{Brambilla:2004jw} relies on the following separation of scales $M \gg Mv \gg Mv^2$ and $M \gg \Lambda_{QCD}$. Here $M$ denotes the heavy quark mass and $v$ the relative quark velocity. Integrating out the physics of the \textit{hard scale} $M$ one arrives at the effective field theory of non-relativistic QCD (NRQCD). Assuming that the extent of the quarkonium is small compared to characteristic scales, one can integrate out also the \textit{soft scale} $Mv$ to obtain potential NRQCD. This EFT describes the evolution of color singlet- and octet wavefunctions interacting with \textit{ultrasoft} gluon fields. The static potential is but one of the non-local Wilson coefficients of this theory. 

The real-time evolution of a static quark-antiquark pair is described by a Wilson loop. If, after time coarse graining, it evolves according to a Schr\"odinger equation we can use it to define the potential (see e.g. \cite{Laine:2006ns})
\begin{align}
i \partial_t W_\Box (t,r) = \Phi(t,r)W_\Box(t,r), \quad V(r) = \lim_{t \rightarrow \infty} \Phi(t,r)\in \mathbb{C}.
\end{align}
Since the real-time Wilson loop is not directly accessible on the lattice one must instead take a detour (see \cite{Rothkopf:2011db,Rothkopf:2009pk}) via extracting the spectral function $\rho_\square$ of the Euclidean time Wilson loop 
\begin{align}
 W_\Box (r,t) = \int d \omega e^{-i \omega t}\rho_\Box (r,\omega)  \leftrightarrow \int d \omega e^{-\omega \tau} \rho_\Box (r,\omega) = W_\Box (r,\tau).\label{eq:specdec}
\end{align}
The inversion of the the relation on the right of \cref{eq:specdec}, based on noisy and sparse input data in general is ill-posed. Ref.~\cite{Burnier:2012az} showed that if a potential picture is applicable $\rho_\square$ will exhibit a skewed Lorentzian peak whose position $\Omega$ encodes ${\rm Re}[V]$ and its width $\Gamma$ encodes ${\rm Im}[V]$. 

The applicability of the potential picture has been established at $T=0$ \cite{Bali:2005fu} (for the $1/M$ corrections see \cite{Koma:2006si}), but it remains an open question whether or in what range it also holds at $T>0$ non-perturbatively. HTL perturbation theory \cite{Laine:2006ns}, which corresponds to a specific $T>0$ scale hierarchy, e.g. predicts a Debye screened ${\rm Re}[V](r)$ with a monotonously increasing ${\rm Im}[V](r)$ (for different hierarchies see e.g. \cite{Brambilla:2008cx}). Previous studies based on quenched lattice QCD \cite{Burnier:2016mxc} and the legacy asqtad action \cite{Burnier:2014ssa,Burnier:2015tda} have observed well-defined skewed Loretzian structures in $\rho_\square$ and have extracted a screened ${\rm Re}[V](r)$. They also found indications for the presence of a sizable ${\rm Im}[V](r)$.

In this proceeding we report on our recent results on the spectral structure of Wilson line correlators in Coulomb gauge obtained from state-of-the-art lattice ensembles with $N_f=2+1$ flavors of light HISQ quarks. In addition we report on preliminary results from newly generated quenched ensembles.

\section{Recent results on HISQ lattices}

In the following we summarize the results on the in-medium potential reported in ref.\cite{Bala:2021fkm}. We utilize high statistics ensembles ($N_{\rm conf}= 2-6 \times 10^4$) based on the HISQ action with $N_f=2+1$ dynamical quark flavors generated by the HotQCD and TUMQCD collaboration \cite{Bazavov:2018wmo,Bazavov:2017dsy,HotQCD:2014kol,Bazavov:2019qoo}. Three grid sizes $N_\sigma^3 \times N_\tau$ are available with $N_\tau = 10,12,16$, among which the ratio $N_\sigma/N_\tau=4$ is kept constant to avoid finite volume artifacts. Temperature is changed via the lattice spacing, spanning a range of $140{\rm MeV}\leq T \leq 2000{\rm MeV}$. For most lattice spacings a $T\approx0$ simulation is available for scale setting. For $T<300$MeV the strange quark mass is set to $m_l/m_s=1/20$ while at higher temperatures we also use $m_l/m_s=1/5$. Wilson loops and Wilson lines in Coulomb gauge were computed on these lattices using the SimulateQCD \cite{mazur2021topological,Bollweg:2021cvl} GPU code.

Our study deploys four different and complementary methods to access spectral information of the Wilson correlators and to extract the position $\Omega(r)$ and width $\Gamma(r)$ of the dominant low-lying spectral structure at different spatial separations $r$. The first method constructs a parametrized model of the spectral function and carries out a $\chi^2$ fit of the parameter values to the Euclidean correlator (see also \cite{Larsen:2019bwy,Larsen:2019zqv}). A study of the cumulants of the imaginary time data revealed \cite{Bala:2021fkm} that only the first and second one have a statistically significant signal-to-noise ratio. If a dominant peak structure is present in $\rho$ the first cumulant contains relevant information on its position (effective mass), the second on its width. As discussed in \cite{Bala:2021fkm}, after we subtract off the UV contribution common to $T=0$ and $T>0$ correlators at small imaginary time, the resulting effective mass exhibits a linear falloff at small $\tau$, which is compatible with the presence of a dominant  Gaussian peak. Amending this structure with a delta peak at smaller and one at higher frequencies constitutes our model to extract $\Omega$ and $\Gamma$.

The second method utilizes the Pad\'e interpolation to reconstruct the spectral functions. After carrying out a discrete Fourier transform on the imaginary time data, the Schlessinger formula is used to obtain a rational interpolation. This interpolation can be decomposed into a series of Breit-Wigner structures and the position and width of the relevant peak can be read off from the pole position closest to the real frequency axis. Mock data tests based on HTL correlators \cite{Burnier:2013fca} have confirmed that the method is able to extract $\Omega$ reliably but underestimates $\Gamma$.

\begin{figure}[t!]
\centering
\includegraphics[scale=0.4]{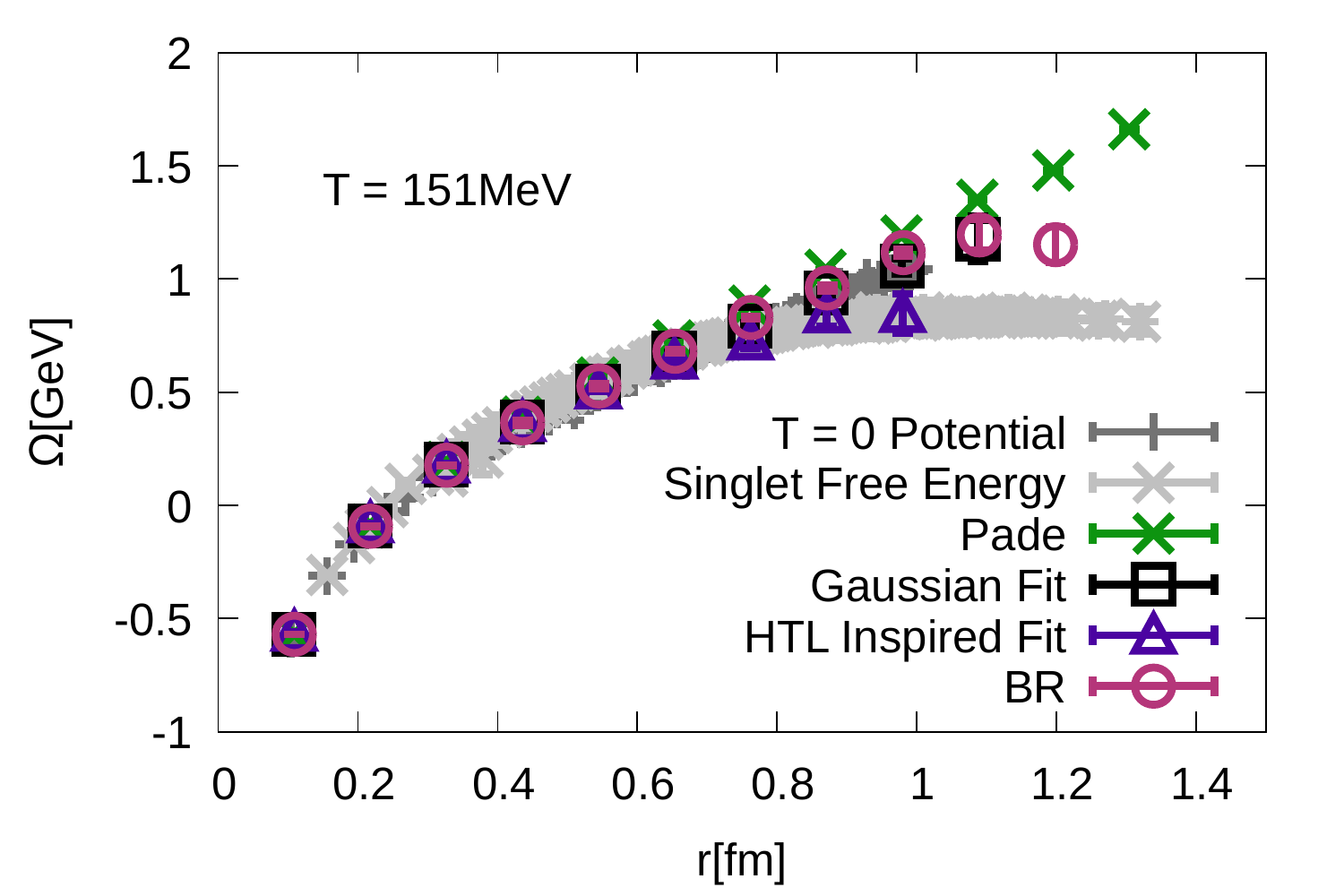}
\includegraphics[scale=0.4]{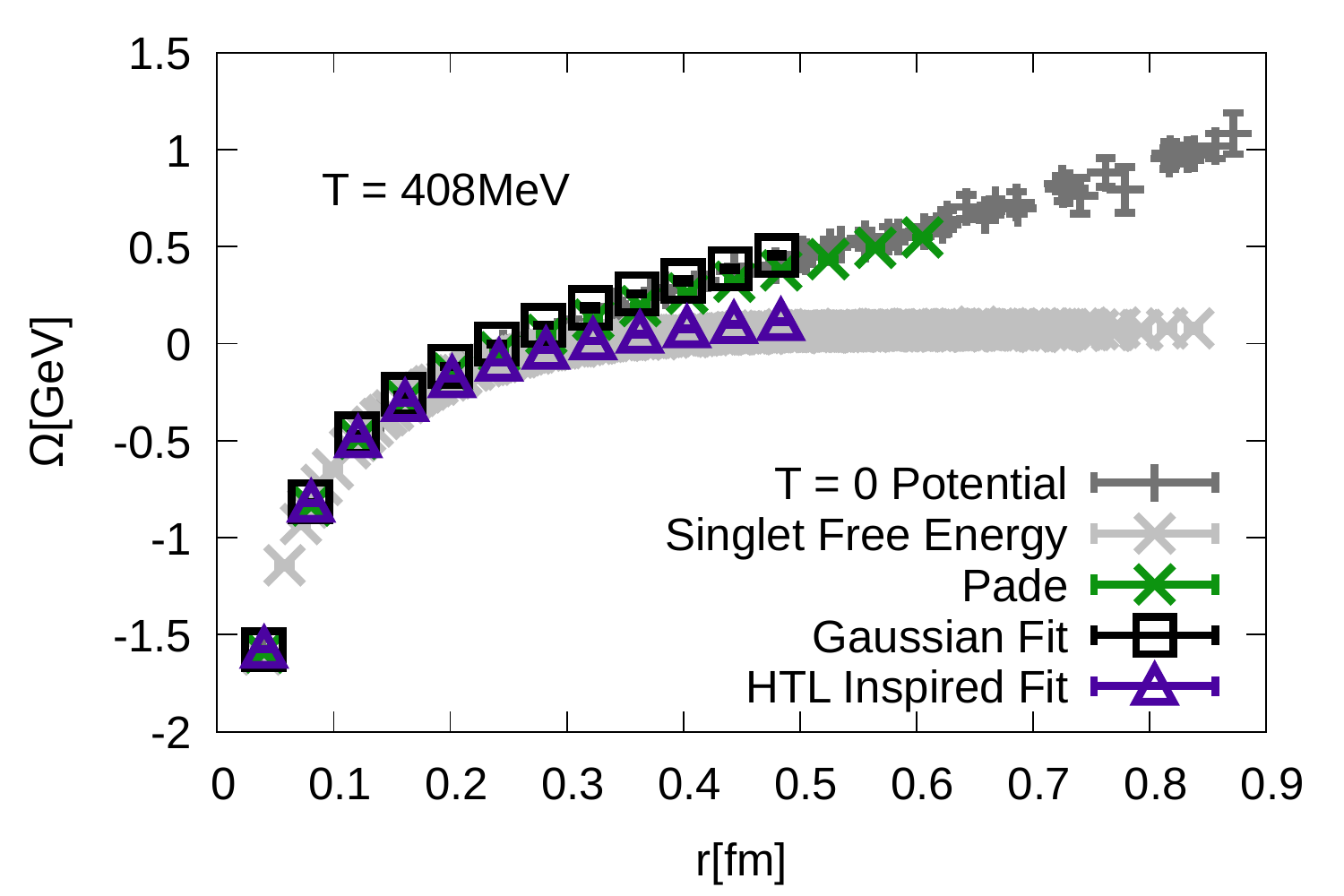}
\caption{Extracted values of $\Omega$ at $T=151$MeV (left) and at $T=408$MeV (right) according to the four complementary methods used in \cite{Bala:2021fkm}. Singlet free energies are shown in light gray and the $T=0$ potential in dark gray.}\label{fig:recReV}\vspace{-0.5cm}
\end{figure}
\begin{figure}[t!]
\centering
\includegraphics[scale=0.4]{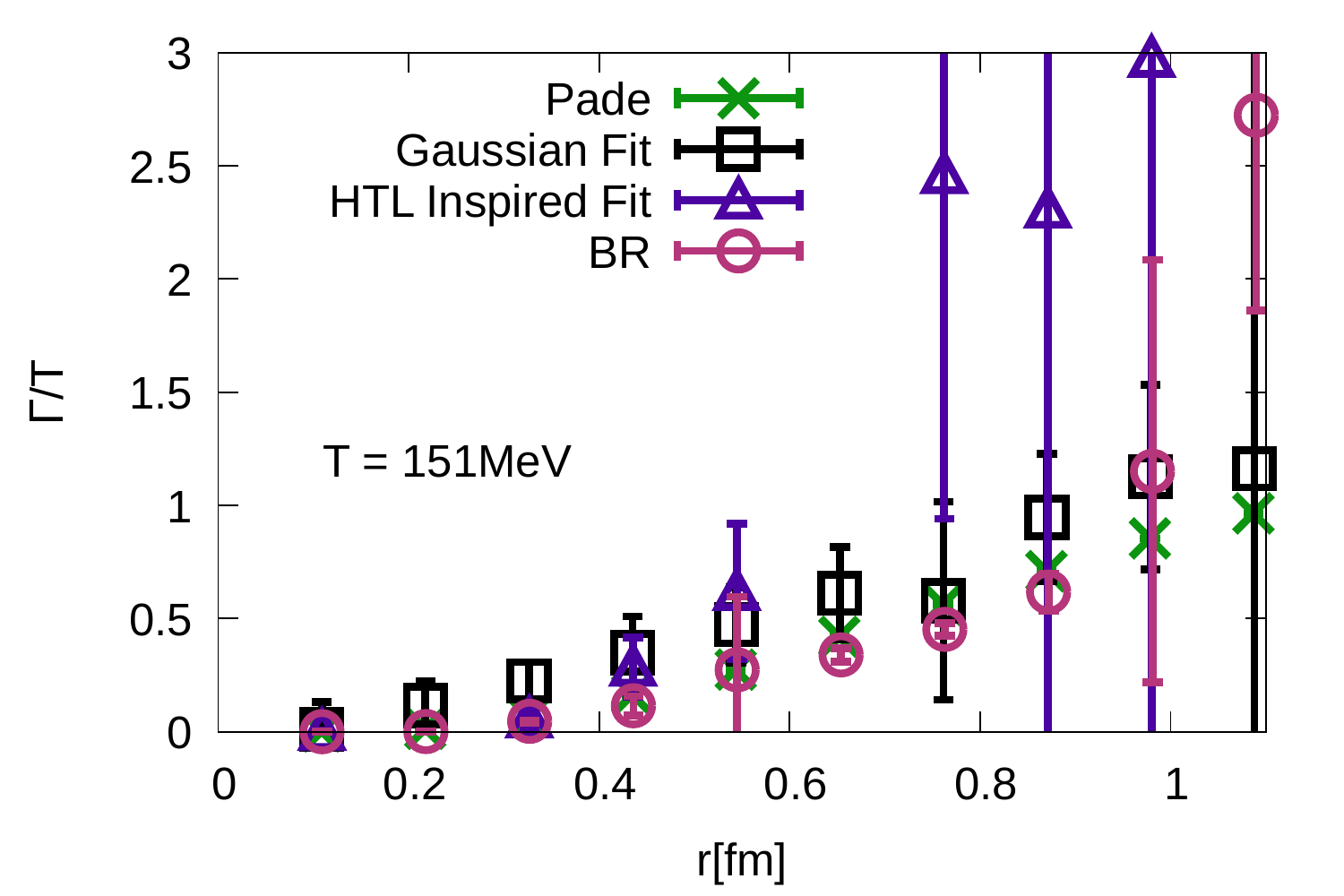}
\includegraphics[scale=0.4]{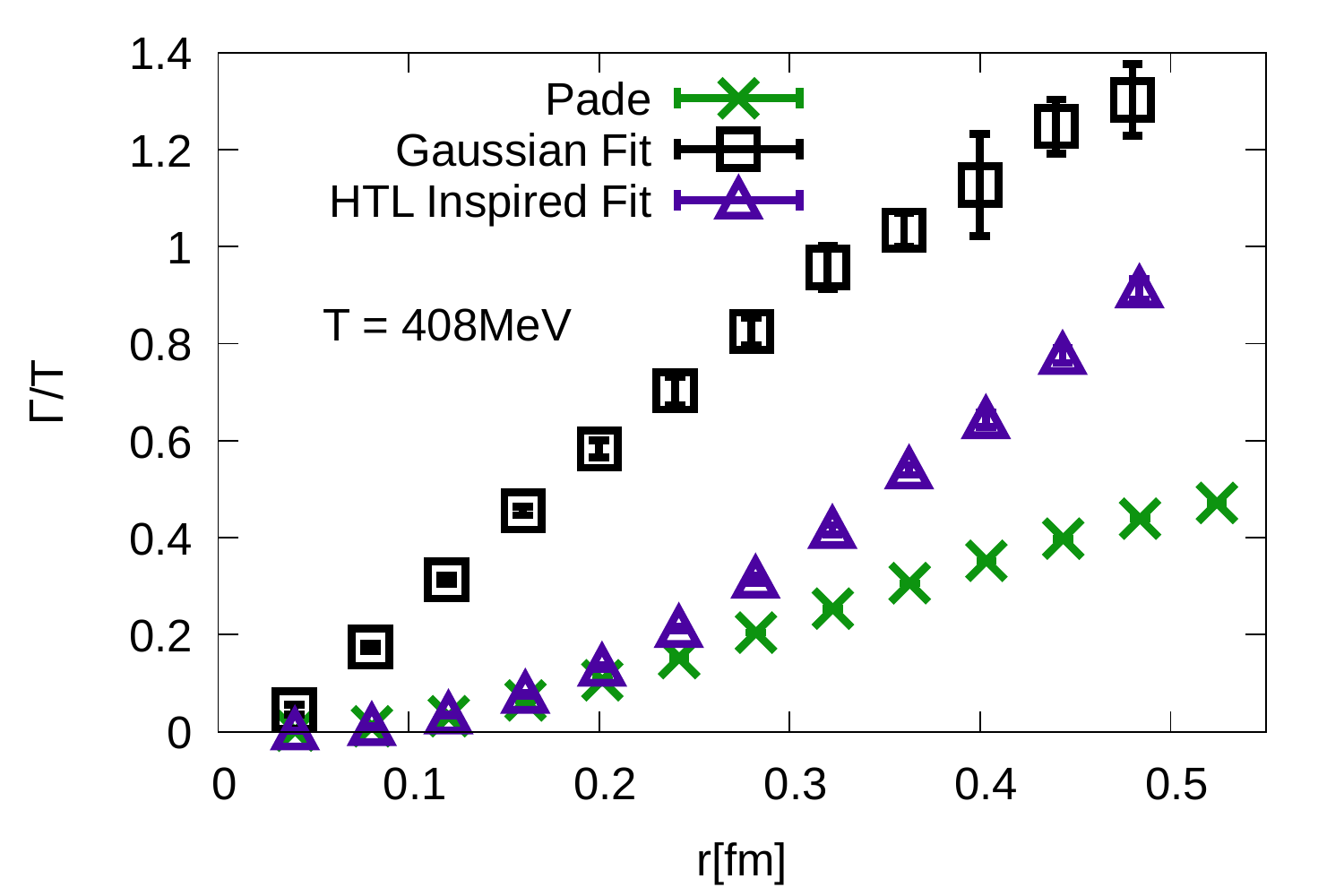}
\caption{Extracted values of $\Gamma$ divided by the temperature at $T=151$MeV (left) and at $T=408$MeV (right) according to the four complementary methods used in \cite{Bala:2021fkm}.}\label{fig:recImV}
\end{figure}

As third method we deployed the Bayesian reconstruction method (BR) \cite{Burnier:2013nla} which has been the workhorse of previous determinations of the in-medium potential in lattice QCD. As by construction is is limited to positive definite spectral functions, we were able to investigate only the ensembles at the lowest available $T$, since at higher temperatures non-positive contributions started to appear in the lattice data, potentially related to the use of the improved HISQ action.

The fourth method \cite{Bala:2019cqu} assumes that the lattice data obeys a specific factorization property that is present in HTL perturbation theory. Similar to the effective mass analysis for well separated stable ground states, it proposes certain ratios of the correlator, which in case that the factorization hypothesis holds exhibit plateaus. The value of the plateau can be related to $\Omega$ and $\Gamma$.

We find that the Gaussian model fits and Pad\'e lead to consistent outcomes. In contrast to previous studies, they however do not show a significant modification of the position of the spectral peak $\Omega(r)$ with temperature. Plotted as black squares and green crosses in \cref{fig:recReV} the two methods agree and the extracted values lie on top of the $T=0$ values of $\Omega(r)$, which is given as dark gray data points. The color singlet free energies are given as light gray points for reference. Only the HTL inspired method (violet triangle) shows signs of an abated rise of $\Omega(r)$ with respect to distance.

\begin{wrapfigure}[36]{r}{0.43\linewidth}
\begin{overpic}[scale=.375,trim= 0 1cm 0 0, clip=true]{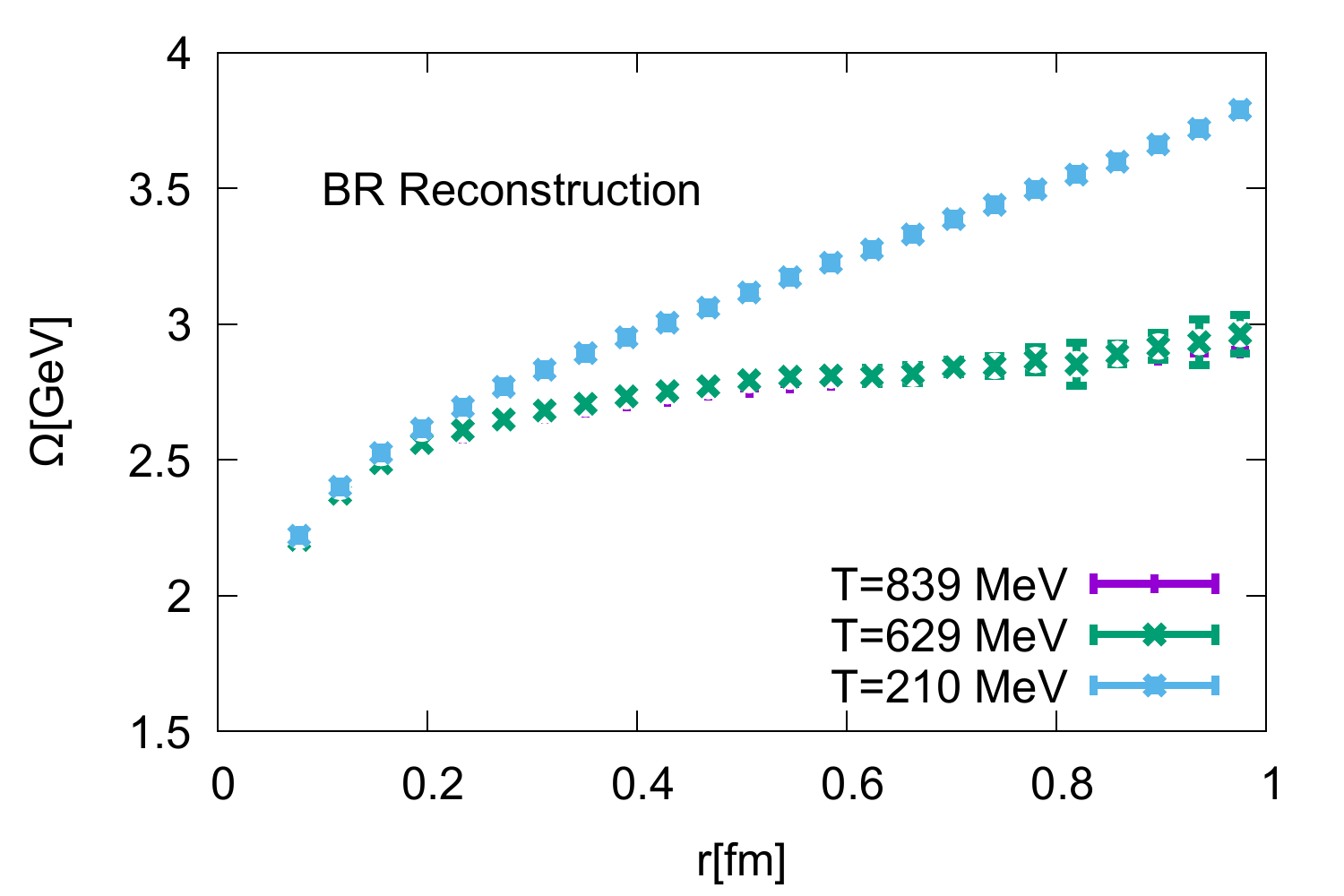}
 \put (25,15) {\color{gray}\textsc{Preliminary}}
\end{overpic}
\begin{overpic}[scale=.375,trim= 0 1cm 0 0, clip=true]{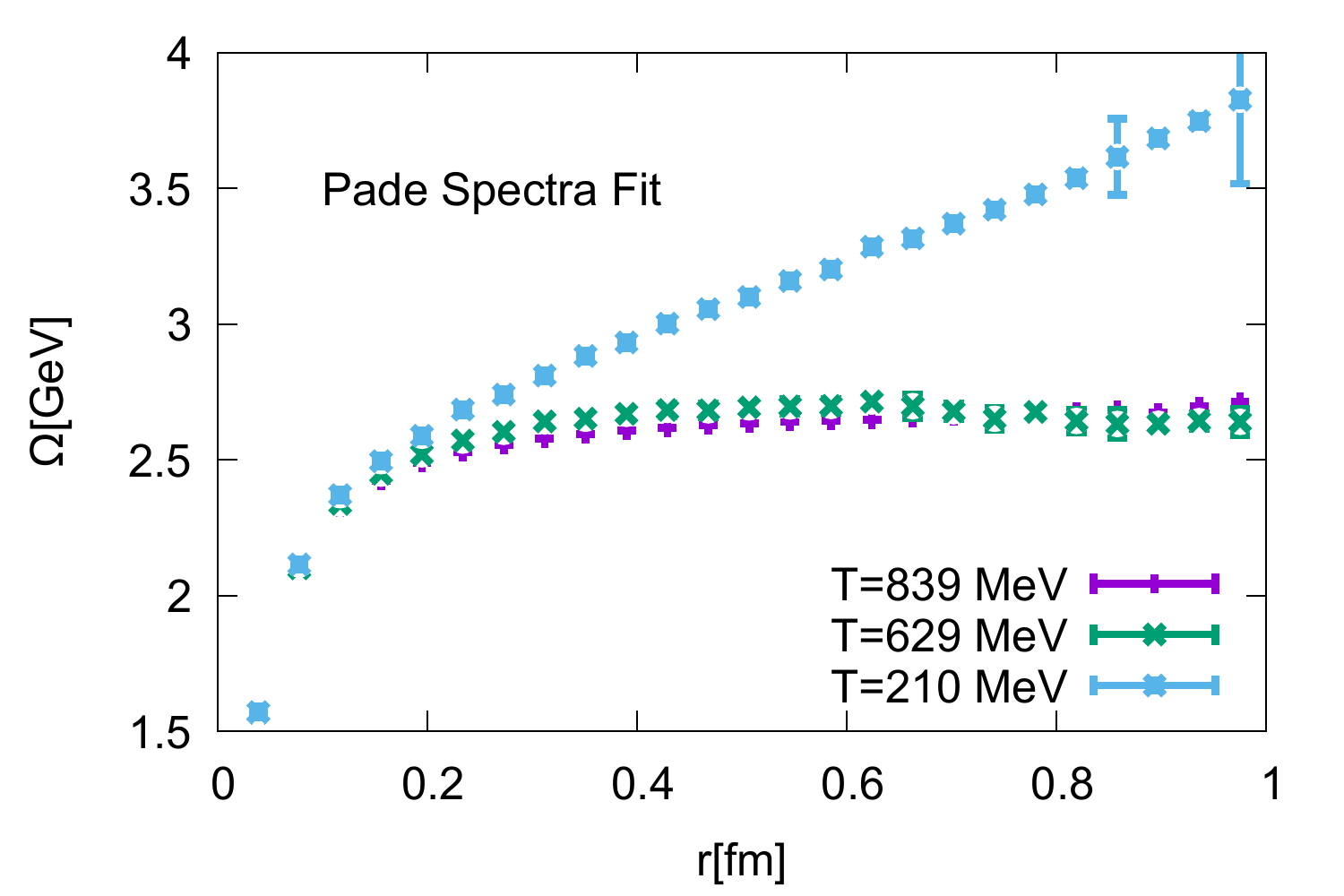}
 \put (25,15) {\color{gray}\textsc{Preliminary}}
\end{overpic}
\begin{overpic}[scale=.375]{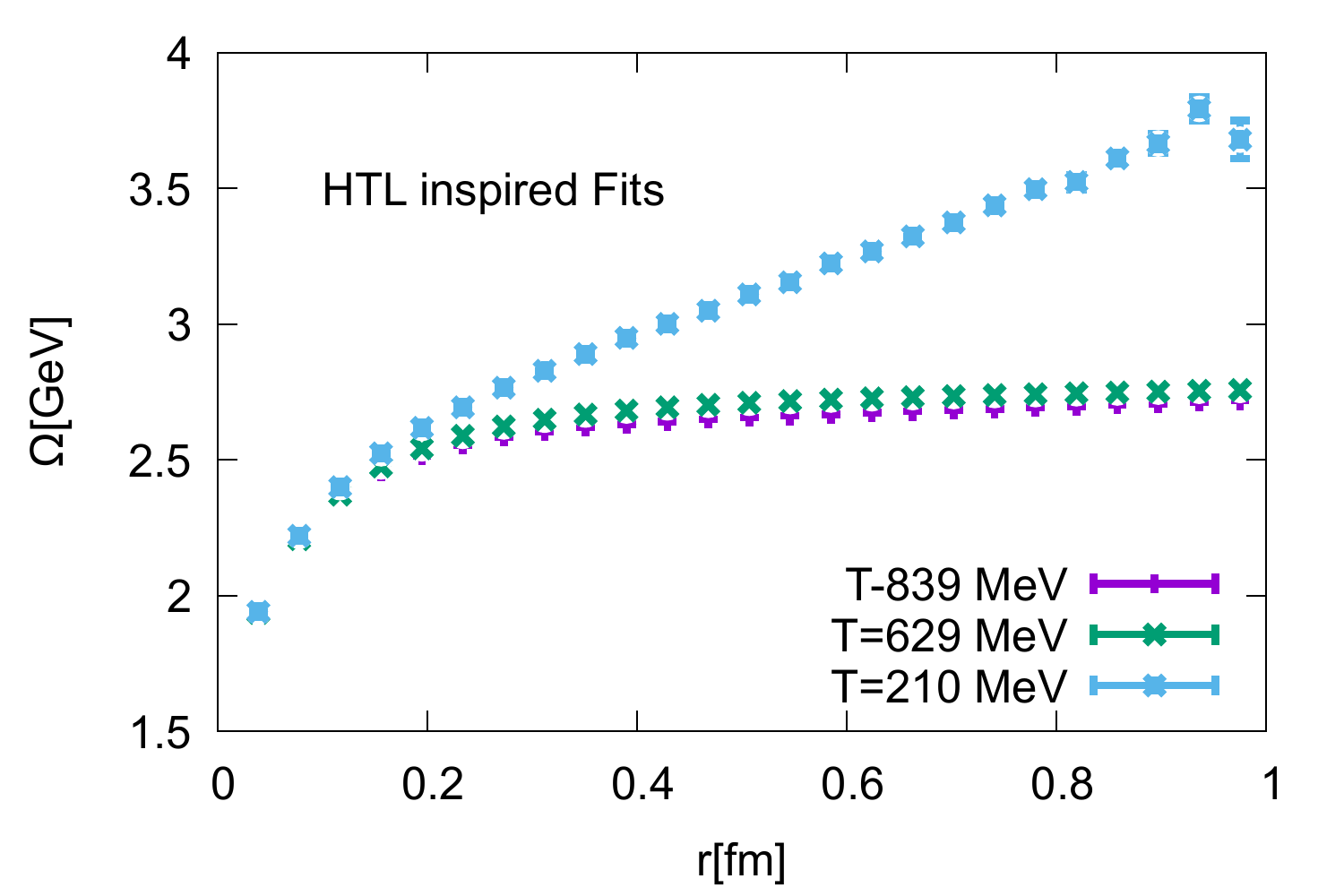}
 \put (25,20) {\color{gray}\textsc{Preliminary}}
\end{overpic}
\caption{Extracted values of the position of the dominant spectral peak $\Omega$ on quenched anisotropic lattices using the BR method (top), the Pad\'e approach (middle) and the HTL inspired fit (bottom). The results from the confined phase at $T=210$MeV are shown as light blue crosses, two results from the deconfined phase at $T=629$MeV and $T=839$MeV as green and violet data points respectively. }
    \label{fig:ReVquenched}
\end{wrapfigure}

What about the width of the spectral structure? In \cref{fig:recImV} we plot $\Gamma/T$ to remove the temperature dependence predicted in HTL perturbation theory. Even at $T=151$MeV (left) most of the methods pick up a finite value of $\Gamma$ above $r=0.4$fm. At $T=408$MeV the onset of a finite $\Gamma/T$ is moved to even smaller distances. While in qualitative agreement, all three methods applicable at this $T$ show significantly different values for $\Gamma/T$ above $r=0.2$fm. It is important to note that the errorbars here include both statistical uncertainty and a rough estimate for the systematic uncertainties of each method (variation of fitting ranges etc.). At different temperatures (see \cite{Bala:2019cqu}) when plotted against the dimensionless product $rT$ the values of $\Gamma/T$ also do not show significant changes.

These results are puzzling, as they differ significantly from previous results in quenched and legacy dynamical lattice QCD. There are two direction for follow up: on the one hand, in order to confirm the results of this HISQ based study, we are generating additional ensembles on finer isotropic lattices to repeat the extraction of spectral features based on correlators with a larger number of points accessible in Euclidean time. On the other hand since the BR method was not applicable at high temperatures and the other three methods had not been used in the extraction of the complex potential in the past, we are generating quenched ensembles with the naive Wilson action, where all methods can be pitted against each other. First preliminary results are reported in the next section.

\section{Preliminary results on quenched lattices}

In order to connect to prior work in the literature, we have generated quenched QCD ensembles on anisotropic grids with the naive Wilson action using the FASTSUM variant of the OpenQCD code \cite{glesaaen_jonas_rylund_2018_2216356}. We choose a spatial grid size of $N_x=64$ and vary temperature at a fixed scale by changing $N_\tau=24,32,40,96,192$. Using $\beta=7$ and bare anisotropy parameter $\xi = 3.5$ results in a physical lattice spacing of $a=0.039$fm and anisotropy $a_\tau/a_x=4$ (used previously in e.g.~\cite{Ikeda:2016czj}). The corresponding set of temperatures reads $T=839,629,503,210,105$MeV. Gauge fixing and computation of Wilson correlators was carried out based on the SIMULATeQCD \cite{mazur2021topological,Bollweg:2021cvl} code.

The first question we wish to address is whether we can qualitatively reproduce prior results on quenched lattices that showed a screened behavior of $\Omega(r)$ in the deconfined phase using all of the methods of \cite{Bala:2021fkm}. For the Wilson action the spectral function of the Wilson loop and Wilson line correlators are positive definite and therefore the BR method can be deployed. As shown in the top panel of \cref{fig:ReVquenched} the BR method reproduces the linearly rising behavior of $\Omega(r)$ in the confined phase. Above the phase transition into the deconfined phase at $T=629$MeV or $T=839$MeV we recover the previous qualitative result of a clear signal for screening.

\begin{wrapfigure}{r}{0.45\linewidth}
       \begin{overpic}[scale=.3]{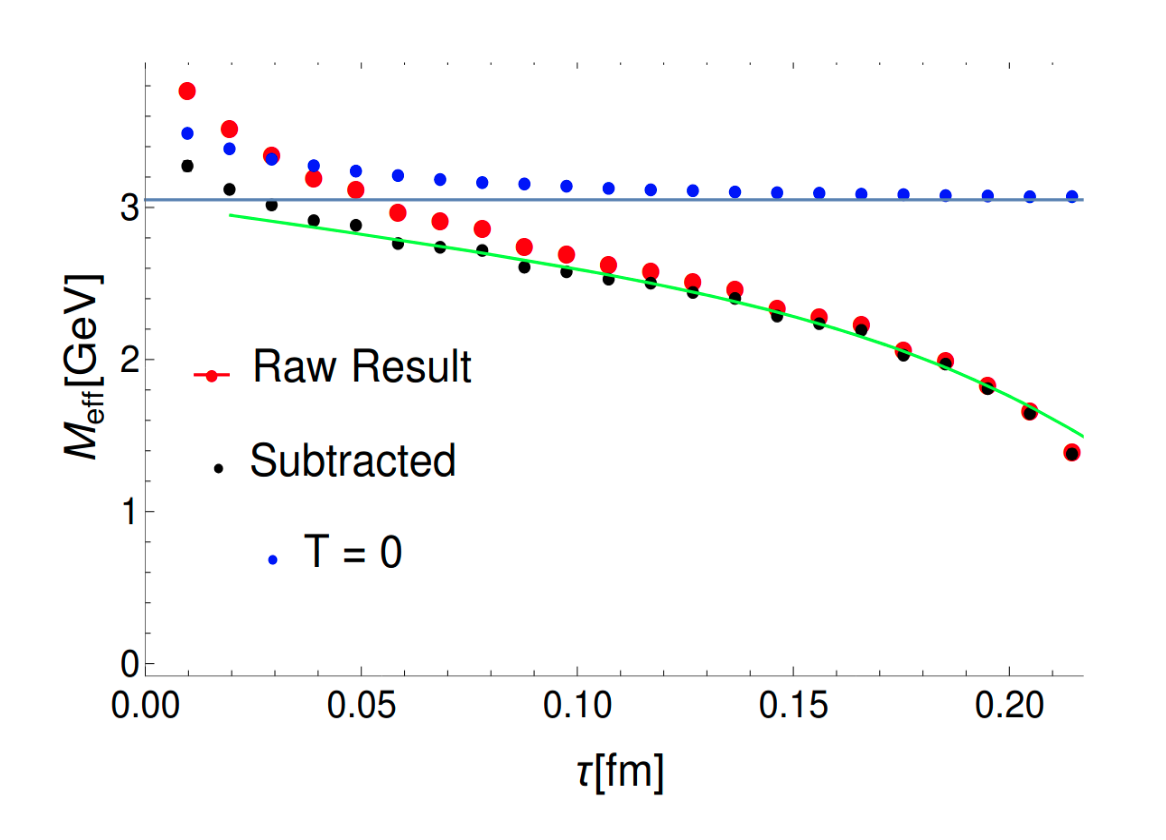}
 \put (45,20) {\color{gray}\textsc{Preliminary}}
\end{overpic}
    \caption{The effective mass $M_{\rm eff}$ of the Wilson line correlator at spatial distance $r/a=12$ at high $T=839$MeV (red circles) and at low $T=210$MeV (blue circles). The effective mass based on the UV subtracted $T>T_C$ correlator is shown as black points.}
    \label{fig:subt_effm}
\end{wrapfigure}
Similarly the Pad\'e method in the middle panel of \cref{fig:ReVquenched} shows a linear rise of $\Omega(r)$ below $T_C$ and an asymptotically flat behavior for $T>T_C$. This is in clear contrast to the result found on the HISQ lattices. The HTL inspired fit, as before, also indicates the presence of a screened behavior at $T>T_C$, as seen in the bottom panel of \cref{fig:ReVquenched}. All methods agree on the slope of the linearly rising $\Omega(r)$ at $T=210$MeV. At $T>T_C$ while in qualitative agreement the BR method appears to show slightly larger values for $\Omega(r)$ than both the Pad\'e and the HTL inspired fit. The errorbars shown here do not yet capture the total error budget.

The reader may ask why we have not shown the Gaussian model fits for the quenched lattices. This model is motivated by the fact that after subtracting the UV contributions to the $T>0$ correlator, the effective masses on HISQ lattices showed a linear behavior towards $\tau\to0$. In particular it was found that the UV behavior of the effective masses is virtually identical below and above the crossover transition. When attempting the subtraction on the quenched anisotropic lattices we found that the UV regime at $T<T_C$ and $T>T_C$ here differs significantly. As shown in \cref{fig:subt_effm}, the effective mass for the Wilson line correlator at $r/a=12$ evaluated at $T=839$MeV (red points) differs from the values at $T=210$MeV (blue points) in the small $\tau<0.04$fm range. Correspondingly $M_{\rm eff}$ based on the UV subtracted correlator (black points) does not show a simple straight line at small $\tau$ rendering the motivation for the Gaussian ansatz moot. 

The difference in the UV behavior can either stem from the fact that the HISQ lattices are isotropic and the quenched ones are anisotropic or alternatively that since quenched QCD exhibits an actual phase transition the continuum below and above $T_C$ is genuinely different. To shed light on this issue we are currently computing the Wilson line correlators also on isotropic quenched lattices (previously used in e.g.~\cite{Burnier:2017bod,Altenkort:2020fgs}), the analysis of which will be reported elsewhere.

\section{Summary}

Our recent study of the position $\Omega$ and width $\Gamma$ of the lowest lying dominant spectral feature in Wilson loop and Wilson line correlators on state-of-the-art HISQ lattices at finite $T$ has revealed an interesting and puzzling result. Several of the complementary methods we used (Gaussian fits, Pad\'e) indicated that the position is independent of temperature while the width of the structure grows with temperature. Only the HTL inspired fits indicated a screened behavior of $\Omega$. In this proceeding we briefly reviewed these findings and presented follow up preliminary results based on anisotropic quenched ensembles. For the Wilson line correlators on these lattices all applicable methods (BR, Pade\'e, HTL inspired fits) show the presence of screening in $\Omega$. In order to conduct further clarifying comparisons, we are currently extending the quenched analysis to isotropic lattices.

\section{Acknowledgments}
The results presented in this proceeding are based upon work supported by the U.S. Department of Energy, Office of Science, Office of Nuclear Physics through the (i) Contract No. DE-SC0012704, and  (ii) Scientific Discovery through Advance Computing (SciDAC) award  Computing the Properties of Matter with Leadership Computing Resources. (iii) R.L., G.P. and A.R. are funded by the Research Council of Norway under the FRIPRO Young Research Talent grant 286883 with computing time provided by the PRACE award on JUWELS at GCS@FZJ, Germany and project NN9578K-QCDrtX from UNINETT Sigma2, Norway . (iv) J.H.W.’s research was funded by the Deutsche Forschungsgemeinschaft (DFG, German Research Foundation) - Projektnummer 417533893/GRK2575 ``Rethinking Quantum Field Theory''. (v) D.B. and O.K. acknowledge support by the Deutsche Forschungsgemeinschaft (DFG, German Research Foundation) through the CRC-TR 211 'Strong-interaction matter under extreme conditions'– project number 315477589 – TRR 211.
\bibliography{references}

\end{document}